\documentclass[letterpaper,reprint,10pt,aps,prb,superscriptaddress,floatfix]
{revtex4-2}
\usepackage[utf8]{inputenc}
\usepackage{bm}
\usepackage[usenames]{color}
\usepackage{multirow}
\usepackage{amssymb}
\usepackage{amsbsy}
\usepackage{amsmath}
\usepackage{appendix}
\usepackage{stmaryrd}
\usepackage[dvipsnames]{xcolor}
\usepackage{graphicx}
\usepackage{epsfig}
\usepackage{placeins}
\usepackage[normalem]{ulem}
\usepackage{bbold}
\usepackage{braket}
\usepackage{blindtext}
\usepackage{soul}

\usepackage{color}

\definecolor{mycolor}{RGB}{0,100,0}

\usepackage{tikz}
\usepackage{natbib}
\usepackage{graphicx}

\usepackage{xcolor}
\definecolor{midnight3}{HTML}{4a6d88}

\usepackage[colorlinks,linkcolor=midnight3,citecolor=midnight3,urlcolor=midnight3]{hyperref}

\definecolor{palette1}{HTML}{A8216B}
\definecolor{palette2}{HTML}{F1184C}
\definecolor{palette3}{HTML}{F36943}
\definecolor{palette4}{HTML}{F7DC66}
\definecolor{palette5}{HTML}{2E9599}

\makeatletter
\newcommand*{\balancecolsandclearpage}{
  \close@column@grid
  \clearpage
  \twocolumngrid
}
\makeatother
\makeatletter
\def\maketitle{
\@author@finish
\title@column\titleblock@produce
\suppressfloats[t]
\let\and\relax
\let\affiliation\@gobble@opt@one
\let\address\affiliation
\let\author\@gobble
\@author@init
\let\@authors\@empty
\let\@authors@curr\@empty
\let\@affil@list\@empty
\let\keywords\@gobble
\let\@keywords\@empty
\let\email\@gobble
\let\@address\@empty
\let\thanks\@gobble
\titlepage@sw{ %
\clearpage
}{}%
}
\makeatother
\usepackage{scalerel}
\usepackage{tikz}
\usepackage{comment}
\usetikzlibrary{calc}
\usetikzlibrary{patterns}
\usetikzlibrary{svg.path}
\definecolor{orcidlogocol}{HTML}{A6CE39}
\tikzset{
  orcidlogo/.pic={
    \fill[orcidlogocol]
svg{M256,128c0,70.7-57.3,128-128,128C57.3,256,0,198.7,0,128C0,57.3,57.3,0,128,
0C198.7,0,256,57.3,256,128z};
    \fill[white] svg{M86.3,186.2H70.9V79.1h15.4v48.4V186.2z}

svg{M108.9,79.1h41.6c39.6,0,57,28.3,57,53.6c0,27.5-21.5,53.6-56.8,53.6h-41.8V79.
1z
M124.3,172.4h24.5c34.9,0,42.9-26.5,42.9-39.7c0-21.5-13.7-39.7-43.7-39.7h-23.
7V172.4z}

svg{M88.7,56.8c0,5.5-4.5,10.1-10.1,10.1c-5.6,0-10.1-4.6-10.1-10.1c0-5.6,4.5-10.1
,10.1-10.1C84.2,46.7,88.7,51.3,88.7,56.8z};
  }
}

\newcommand\orcid[1]{\!%
  \href{https://orcid.org/#1}{%
    \mbox{%
      \scaleto{%
        \begin{tikzpicture}[yscale=-1,transform shape]
          \pic{orcidlogo};
        \end{tikzpicture}
      }{8pt}%
    }%
  }%
}
\begin{document}

\title{Boundary-driven magnetization transport in the spin-$1/2$
XXZ chain:\\ Role of the system-bath coupling strength and timescales}

\author{Mariel Kempa~\orcid{0009-0006-0862-4223}}
\email{makempa@uos.de}
\affiliation{University of Osnabr{\"u}ck, Department of Mathematics/Computer
Science/Physics, D-49076 Osnabr{\"u}ck, Germany}

\author{Markus Kraft~\orcid{0009-0008-4711-5549}}
\affiliation{University of Osnabr{\"u}ck, Department of Mathematics/Computer
Science/Physics, D-49076 Osnabr{\"u}ck, Germany}

\author{Sourav Nandy~\orcid{0000-0002-0407-3157}}
\affiliation{Max Planck Institute for the Physics of Complex Systems, D-01187
Dresden, Germany}

\author{Jacek Herbrych~\orcid{0000-0001-9860-2146}}
\affiliation{Wroclaw University of Science and Technology, 50-370 Wroclaw,
Poland}

\author{Jiaozi Wang~\orcid{0000-0001-6308-1950}}
\affiliation{University of Osnabr{\"u}ck, Department of Mathematics/Computer
Science/Physics, D-49076 Osnabr{\"u}ck, Germany}

\author{Jochen Gemmer~\orcid{0000-0002-4264-8548}}
\affiliation{University of Osnabr{\"u}ck, Department of Mathematics/Computer
Science/Physics, D-49076 Osnabr{\"u}ck, Germany}

\author{Robin Steinigeweg~\orcid{0000-0003-0608-0884}}
\email{rsteinig@uos.de}
\affiliation{University of Osnabr{\"u}ck, Department of Mathematics/Computer
Science/Physics, D-49076 Osnabr{\"u}ck, Germany}

\date{\today}

%-------------------------------------------------------------------------------
% Abstract
%-------------------------------------------------------------------------------

\begin{abstract}
{Understanding the transport properties of quantum many-body systems is a
central
challenge in condensed matter and statistical physics. Theoretical studies
usually rely on two main approaches: Dynamics of linear-response functions in
closed systems and boundary-driven dynamics governed by Markovian master equations
for open systems. While the equivalence of their dynamical behavior has been explored
in recent studies, a systematic comparison of the transport coefficients obtained from
these two classes of approaches remains a largely open question.
Here, we address this gap by comparing and contrasting the dc diffusion constant
$\mathcal{D}_{\text{dc}}$ according to the two approaches, focusing on the specific
example of magnetization transport in the spin-$1/2$ XXZ chain. Using exact numerical
simulations for finite system sizes, we find (i) a clear mismatch between the two
$\mathcal{D}_{\text{dc}}$ and (ii) a strong dependence of $\mathcal{D}_{\text{dc}}$
on the system-bath coupling strength for the open system, where neither (i) nor (ii) tend
to vanish in the thermodynamic limit.
These findings suggest limitations of the open-system approach to transport coefficients.
To gain insight into the origin of (i) and (ii),  we go beyond $\mathcal{D}_{\text{dc}}$
and analyze the full time dependence of the diffusion coefficient $D(t)$ in the open system.
In this way, we find that both (i) and (ii) vanish up to a finite time scale. While this
time scale gradually increases with system size and tends to be macroscopic in the
thermodynamic limit, this increase is still slow compared to the increase of the time to
reach the steady state, where (i) and (ii) do not vanish. This observation can be seen as
a wrong, yet unavoidable order of limits of long times first and large system sizes
afterwards.}

\end{abstract}

\maketitle

%------------------------------------------------------------------------------
% Introduction
%------------------------------------------------------------------------------
\section{Introduction}
Over the last few decades, studying nonequilibrium dynamics in quantum many-body systems has become a key area of research in both theoretical and experimental condensed matter physics, with broad ramifications ranging from fundamental questions in statistical mechanics to advanced applications in quantum control and quantum computation \cite{Polkovnikov_rev, Abanin_rev, Bloch_RMP, Monroe_rev}. One of the most ubiquitous manifestations of nonequilibrium behavior are transport processes of conserved quantities such as energy, particle number and magnetization, which are relevant for both closed/isolated and open quantum systems coupled to external environments. A comprehensive understanding of transport is thus crucial for elucidating the emergent macroscopic behavior as well as the underlying microscopic mechanisms governing quantum many-body dynamics such as the build up of nonequilibrium steady states (NESS) and the relaxation to them \cite{Bertini_rev}. Over the past years, notable progress in this field has been driven by the development of new theoretical concepts like eigenstate thermalization hypotheses \cite{ETH_srdnicki_1994, ETH_Deustch, Deutsch_2018, Riemann_ETH} and dynamical quantum typicality \cite{Gemmer_typicality, Popescu2006, Goldstein_typicality, Riemann_Typicality, Heitmann_typicality}, as well as sophisticated analytical and numerical methods and advanced experimental techniques. We now provide a brief overview of some of the key numerical and analytical methods commonly employed to study such problems, which are particularly relevant to the present work.

\par 
In an isolated or closed system, transport phenomena are often studied within the framework of linear response theory (LRT) leading to a description of the system's dynamics in terms of equilibrium correlation functions. Loosely speaking, such correlation functions describe the response of the system close to equilibrium due to an external perturbation. Among the foundational results of this formalism is the well-known Kubo formula \cite{Green_LRT, Kubo, Zwanzig_LRT, forster75}, which provides a quantitative link between the microscopic dynamics and macroscopic transport coefficients. Despite their formal clarity, the concrete analytical or numerical calculation of correlation functions for specific interacting quantum many-body models often remains a subtle task. Even for systems with seemingly simple interactions, such as the Bethe-Ansatz integrable spin-$1/2$ XXZ chain or one-dimensional Hubbard models \cite{Eckle, Takahashi:1999bgb}, the calculation of response functions which involve matrix elements of some suitable (usually) local operator between different eigenstates, have turned out to be notoriously difficult for all practical purposes. Although, some breakthroughs in these direction have indeed been made by generalized hydrodynamics (GHD) \cite{GHD_00, GHD_0, GHD_1, GHD_2, GHD_3}, such formalism are (strictly speaking) tailor made for integrable systems. Therefore, several realistic and experimentally relevant questions in the presence of integrability breaking generally remain outside the scope of the GHD formalism.
\\\\
\par 
On the numerical frontier, an alternative to traditional LRT-based closed-system technique is given within the framework of boundary-driven open quantum systems. In such a setup, the system is coupled to an environment, for instance, modeled by external baths at different temperatures or chemical potentials \cite{Gemmer_boundary_drive, Prosen_2009, Znidaric_2011_PRE, Znidaric_PRL_2011, Landi_rev}. This coupling drives the system out of equilibrium and induces transport, typically resulting in a NESS with a constant current in the bulk and a characteristic density profile in the long-time limit. A widely employed approach for describing the dynamics of such open-system scenarios is provided by Gorini–Kossakowski–Sudarshan–Lindblad (GKSL) quantum master equation \cite{Gorini_master_equation, Lindblad1976}. While the derivation of the GKSL master equation from a microscopic system-bath model is a nontrivial task \cite{Wichterich2007, DeRaedt2017}, it provides the most general form of a time-local (Markovian) master equation whose dynamical map is completely positive and trace-preserving (CPTP), thus ensuring the evolution of any physical density matrix into another valid density matrix \cite{Breuer2007}. This formalism offers considerable advantages for numerical simulations, particularly in conjunction with stochastic methods or matrix-product-state techniques based upon vectorized density matrices.
\par
In light of the preceding discussion on methodological approaches to study transport, it becomes imperative to examine the correspondence between the frameworks based on closed and open quantum systems. Remarkably, despite the fundamental importance of this question, especially in the context of benchmarking state-of-the-art numerical methods for transport, it has attracted serious attention only in recent years \cite{Dynamic_corrspondence_0, Dynamic_corrspondence_1, Dynamic_corrspondence_2}. While latest works by some of the authors of this paper have unveiled, in certain cases, a clear connection between the dynamics of the density-profile build up in the closed and open quantum systems, it remains yet unclear how such a correspondence is reflected within transport coefficients. In this paper, we essentially attempt to bridge this gap, by studying the high-temperature (infinite-temperature in principle) magnetization transport in the boundary-driven spin-$1/2$ XXZ chain as a concrete example. 
\par 
%------------------------------------------------------------------------------
\begin{figure}[t]
\includegraphics[width=1.\columnwidth]{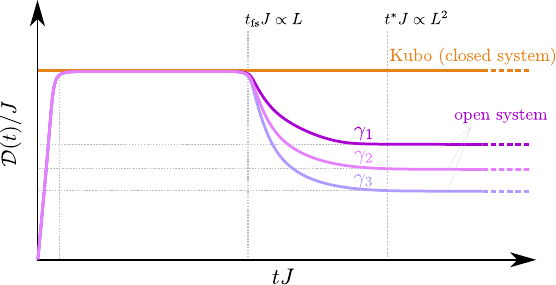}
\caption{Sketch summarizing our key findings for the time-dependent
diffusion constant $\mathcal{D}(t)$ from the open-system approach (purple). The
closed-system Kubo-based approach \cite{Karrasch2014,kraft2024,Wang2024} is
indicated (orange). An early-time plateau shows a (nearly) $\gamma$-independent
agreement up to a system-size–dependent scale $t_{\mathrm{fs}}J \propto L$. At
long times, the NESS-based dc values disagree with Kubo and strongly depend on
the system–bath coupling $\gamma$.}
\label{sketch}
\end{figure}
%------------------------------------------------------------------------------

Using exact numerical simulations for finite system sizes, we find (i) a clear mismatch between the two
$\mathcal{D}_{\text{dc}}$ and (ii) a strong dependence of $\mathcal{D}_{\text{dc}}$
on the system-bath coupling strength for the open system, where neither (i) nor (ii) tend
to vanish in the thermodynamic limit. These findings suggest limitations of the open-system approach to transport coefficients.
To gain insight into the origin of (i) and (ii),  we go beyond $\mathcal{D}_{\text{dc}}$ and analyze the full time dependence of
the diffusion coefficient $D(t)$ in the open system, as sketched in Fig.\ref{sketch}. In this way, we find that both (i) and (ii) vanish up to a finite time scale.
While this time scale gradually increases with system size and tends to be
macroscopic in the thermodynamic limit, this increase is still slow compared to
the increase of the time to reach the steady state, where (i) and (ii) do not
vanish. This observation can be seen as a wrong, yet unavoidable order of limits
of long times first and large system sizes afterwards.

\par 
The rest of the paper is organized as follows. In Sec.~\ref{sec: model}, we introduce the models for which we perform our numerical analysis. In Sec.~\ref{sec:key_ideas}, we present a comprehensive discussion of some of the key concepts related to this paper, namely the basic ideas of boundary-driven techniques and the extraction of diffusion constants from both, the boundary-driven open-system based method and from closed-system dynamics. In Sec.~\ref{sec:methods}, we discuss the numerical methods we use for this work. This is followed by Sec.~\ref{sec: results} where we present our main numerical results and, finally, we present our conclusion in Sec.~\ref{sec: conclusions}.

%------------------------------------------------------------------------------
% model
%------------------------------------------------------------------------------
\section{model} \label{sec: model}
As a concrete example, we consider the one-dimensional spin-$1/2$ XXZ model described
by the Hamiltonian
\begin{equation}\label{eq: integrable system}
H = J \sum_{r=1}^{L-1} \left[ S^x_r S^x_{r+1} + S^y_r S^y_{r+1} + \Delta S^z_r
S^z_{r+1} \right],
\end{equation}
where $S^{x,y,z}_r$ are spin-$1/2$ operators at site $r$, $J>0$ is
the antiferromagnetic coupling constant and $\Delta$ denotes the anisotropy in
$z$ direction. We note that the sign of $J$ for this work does not really matter since we shall be focusing on infinite-temperature physics. The total number of sites is denoted by $L$ and throughout our work we focus on open boundary conditions.
In this system the global magnetization is conserved, i.e.~$[H,\sum_rS_r^z]=0$. 
Note that this global conservation of the total spin essentially makes the study of spin transport a meaningful and well-defined question.
It is well-known that $H$ is integrable in terms of the Bethe Ansatz for any 
value of $\Delta$ and exhibits rich transport properties at high temperatures 
ranging from ballistic ($\Delta < 1$) through superdiffusive ($\Delta = 1$) to
diffusive ($ \Delta > 1$) transport behavior \cite{Bertini_rev, Gopalakrishnan_2023, Bulchandani_2021}.
For our further investigations, we focus on $\Delta \in \{1.5, 3.0\}$, corresponding to the
easy-axis regime where spin transport in the integrable system becomes
diffusive. Furthermore, we also consider a second system $H'$ with
\begin{equation}\label{eq: nonintegrable system}
H' = H + J \sum_{r=1}^{L-2} \Delta' S^z_r S^z_{r+2}\ ,
\end{equation}
where the additional next-to-nearest neighbor interactions of strength 
$\Delta'>0$ breaks the integrability of the system and the model becomes
more generic.

%------------------------------------------------------------------------------
% open system scenario
%------------------------------------------------------------------------------
\section{Key Background Concepts}
\label{sec:key_ideas}
In this section, we introduce the underlying boundary-driven system. Additionally, we outline two key background concepts relevant for extracting diffusion constants from both boundary-driven open systems and closed systems.
%
%------------------------------------------------------------------------------
\begin{figure}[b]
\includegraphics[width=0.85\columnwidth]{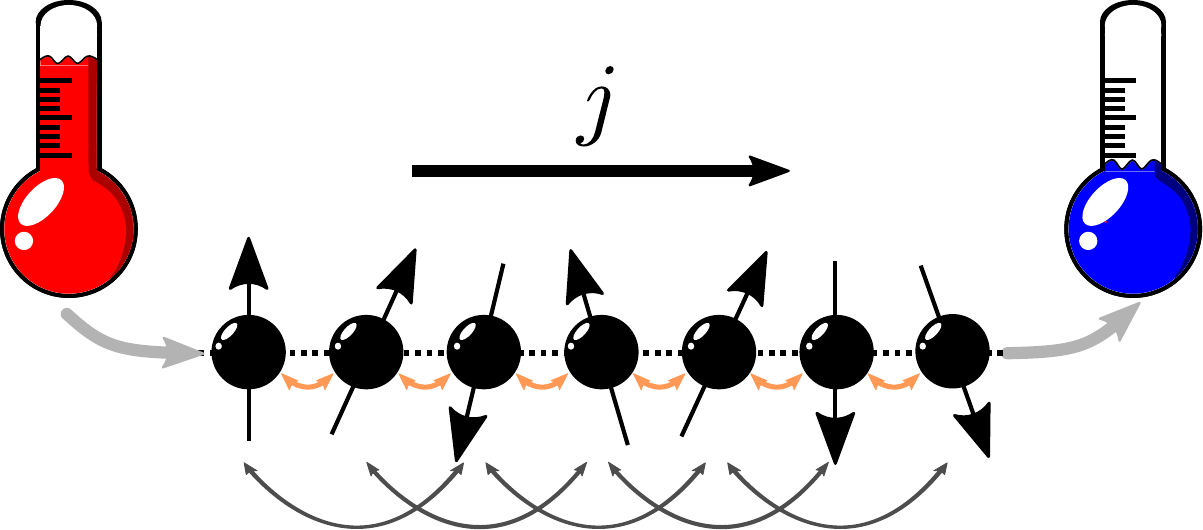}
\caption{Sketch of the open quantum system scenario, where a spin-$1/2$ XXZ chain with open boundaries is coupled at its ends to two Lindblad baths. Thus, transport is induced and at sufficient long times the nonequilibrium steady state is reached featuring a constant current $j_{\text{ness}}$ and a characteristic density profile. Since we consider an integrable and nonintegrable scenario, corresponding nearest- and next-to-nearest neighbor interactions are indicated.}
\label{Fig1}
\end{figure}
%------------------------------------------------------------------------------
%
\subsection{The boundary-driven system: Set-up and basics}
\label{boundary_drive}
As mentioned previously, in this work we consider an open-system scenario, where we couple our model to an environment as sketched in Fig.~\ref{Fig1}.
It is convenient to describe such scenarios in terms of the most general form of a Markovian quantum dynamics governed by the
GKSL master equation of the form \cite{Lindblad1976, Gorini_master_equation}
\begin{equation}
    \dot{\rho}(t) = \mathcal{L}\rho(t) = -i[H, \rho(t)] + \mathcal{D}\rho(t)\ .
\label{eq:lindblad}
\end{equation}
The first term on the r.h.s.\ of Eq.\,(\ref{eq:lindblad}) models the coherent (unitary) dynamics 
of the closed system $H$, while the latter part effectively describes the 
decohering and dissipative influence of the environment.
In general, the damping term reads
\begin{equation}
      \mathcal{D}\rho(t) = \sum_j \alpha_j \left(
          L_j \rho(t) L_j^\dagger - \frac{1}{2} \{ \rho(t), L_j^\dagger L_j \}
        \right),
    \label{eq:dissipator}
    \end{equation}
where $\{\cdot,\cdot\}$ is the anticommutator. Moreover, $\alpha_j$ and $L_j$ 
denote non-negative rates and the Lindblad jump operators, respectively. 
Here, we consider the jump operators of the following form \cite{Bertini_rev}:
\begin{equation}\label{eq: Lindblad operators}
\begin{aligned}
    &L_1 = S^+_{1}, \qquad           &&\alpha_1 = \gamma(1+\mu), \\
    &L_2 = L_1^\dagger = S^-_{1}, \qquad &&\alpha_2 = \gamma(1-\mu), \\
    &L_3 = S^+_{L}, \qquad           &&\alpha_3 = \gamma(1-\mu), \\
    &L_4 = L_3^\dagger = S^-_{L}, \qquad &&\alpha_4 = \gamma(1+\mu) 
\end{aligned}
\end{equation}
where $\gamma$ is the system-bath coupling strength and $\mu$ is the driving strength.
%
%
%------------------------------------------------------------------------------
\begin{figure}[t]
\includegraphics[width=0.85\columnwidth]{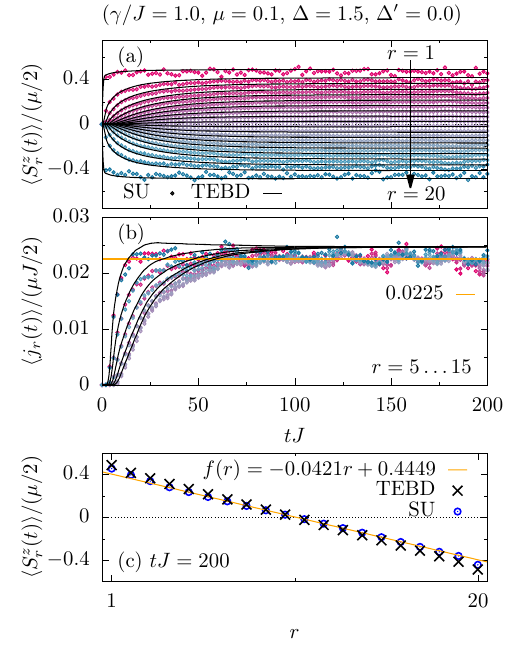}
\caption{Time evolution in the open system for the integrable spin-1/2 XXZ chain
with anisotropy $\Delta = 1.5$ and $L = 20$ sites, as resulting for strong
system-bath coupling $\gamma/J = 1.0$ and weak driving $\mu = 0.1$. Numerical
data are obtained from exact stochastic unraveling (SU) with $N_{\text{traj}} = 50000$
trajectories and time-evolving block decimation (TEBD) with bond dimension $\chi = 150$. (a) Local densities, (b) local currents, and (c) steady-state
profile at time $t J = 200$. In (b) the average over the time window $tJ=100\dots 200$ for the shown bulk sites of the SU data is indicated by the yellow line. Panel (c) depicts the linear function $f(r)$ from fitting over bulk sites $r=5..15$ of SU data at times $tJ=200$.}
\label{Fig2}
\end{figure}
%------------------------------------------------------------------------------
It is straightforward to see from Eq.\,(\ref{eq: Lindblad operators}) that the Lindblad jump operators model baths which, in the absence of Hamiltonian, i.e.~for the case of single-site driving, induce a net magnetization $\mu$ at the left edge (first site) and $-\mu$ at the right edge (last site) \cite{Bertini_rev}. Such a set-up leads to a NESS with position-dependent magnetization along the chain and a non-zero time-independent current $j_{\text{ness}}$ for a finite system. Note that, for $\mu = 0$, one has a trivial steady state $\rho \sim \mathbb{1}$, i.e.~an infinite-temperature state, and one can interpret Eq.\,(\ref{eq: Lindblad operators}) as spin driving at infinite temperature. For non-zero $\mu$, the NESS current is finite and is the main observable \cite{Bertini_rev}. Also, note that the linear response regime demands that $\mu$ should be chosen such that the NESS current $j_{\text{ness}}/\mu$ is independent of $\mu$, leading to desirable values of $\mu \ll 1$.

\par 
The GKSL master equation is hard to solve numerically for a reasonable system size $L$ since the complexity grows as $d^{2L}$, where $d$ is the dimension of the local Hilbert space. To access large system sizes beyond the scope of exact diagonalization, efficient and powerful techniques are employed to find well approximated solutions, as discussed in Sec.~\ref{sec:methods}.

%------------------------------------------------------------------------------
% diffusion constant
%------------------------------------------------------------------------------
\subsection{Extracting diffusion constants from boundary-driven systems}
Now, we discuss how to extract the spin diffusion constant, the central quantity we probe in the present work. To this end, we note that the magnetization at site $r$, at time $t$ is given by 
\begin{equation}
    \langle S^z_r(t) \rangle = \operatorname{Tr}[\rho(t) S^z_r]\ .
\label{eq: local magnetization}
\end{equation}
The time-dependent local spin current (for both $H$ and $H^{\prime}$) is given by
\begin{equation}
\langle j_r(t) \rangle = \operatorname{Tr}[\rho(t) j_r]\ ,\ \ \
j_r=J(S_r^xS_{r+1}^y - S_r^yS_{r+1}^x)\ .
\label{eq: local current}
\end{equation}
The expression for the current in Eq.\,(\ref{eq: local current}) can be derived from the lattice version of the continuity equation for the global conservation of the total spin i.e.~, $[H, \sum_{r}S_{r}^{z}]=0$, as discussed in \cite{Bertini_rev}. 

%--------------- ---------------------------------------------------------------
\begin{figure}[b]
\includegraphics[width=0.85\columnwidth]{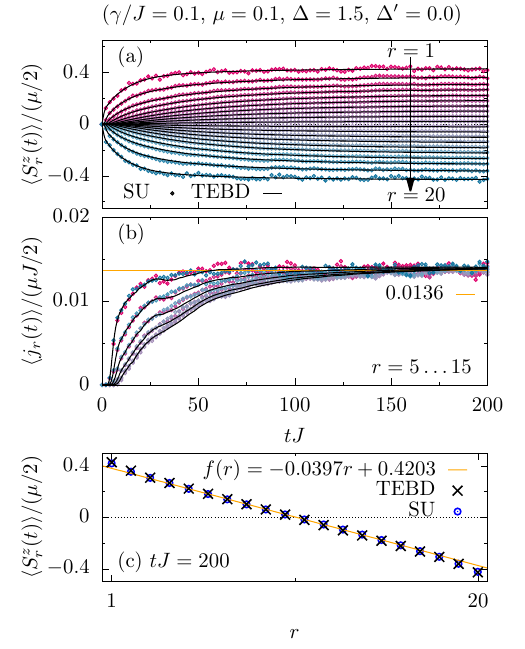}
\caption{Similar data as the one in Fig.\ \ref{Fig2}, but now for a
small system-bath coupling strength $\gamma/J = 0.1$.}
\label{Fig3}
\end{figure}
%------------------------------------------------------------------------------

Given the local magnetization and the local current, the site- and
time-dependent diffusion constant is defined as (according to Fick's law)
\begin{equation} \label{eq: diffusion_constant}
    \mathcal{D}_r(t) = -\frac{\langle j_r(t)\rangle}{\langle \nabla
S_r^z(t)\rangle} \, , \quad \nabla
S_r^z = S_{r+1}^z - S_r^z\ ,
\end{equation}
with $t > 0$. Note that, at $t = 0$, $\mathcal{D}_r(t)$
is not defined, due to $\langle j_r(0)\rangle = 0$ and $\langle \nabla
S_r^z(0)\rangle = 0$. However, as we will see later, $\lim_{t \to 0}
\mathcal{D}_r(t) = 0$. 

In a similar way, one can extract the diffusion constant in the NESS, which we
dub as the dc diffusion constant,  since this is the result for the $t \rightarrow
\infty$ (or $\omega \rightarrow 0$) limit. Note that, in the NESS, the current
is independent of space coordinate $r$, not only in the
bulk (by the very definition of NESS). Assuming that the density profile in
the NESS is linear in the bulk, we thus obtain

\begin{equation}\label{eq: diffusion constant from NESS}
    \mathcal{D}_{\text{dc}} = -\frac{j_{\text{ness}}}{\nabla S_{\text{ness}}^z}\ .
\end{equation}
The above assumption of a linear density profile applies
to all parameters studied later, and deviations occur only in
the close vicinity of the bath contact.

%
%------------------------------------------------------------------------------
\begin{figure}[t]
\includegraphics[width=0.85\columnwidth]{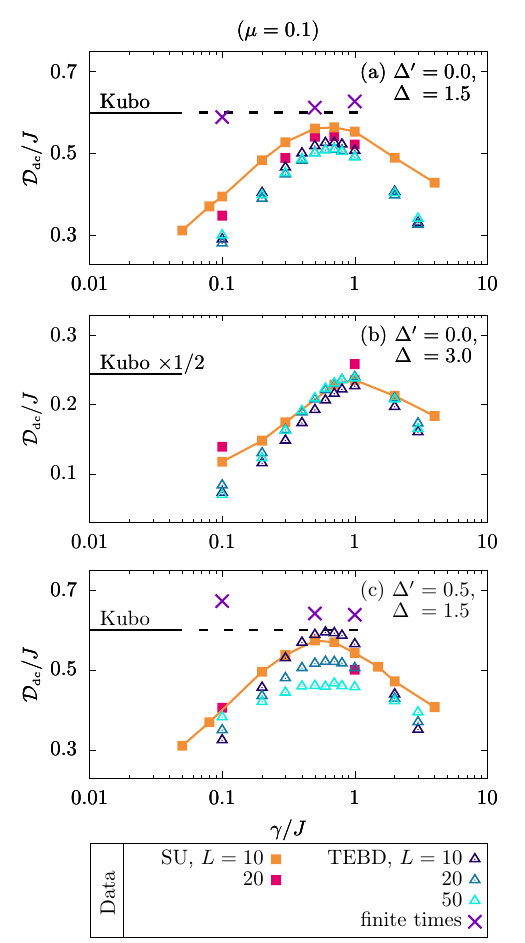}
\caption{Open-system diffusion constant $\mathcal{D}_{\text{dc}}$ versus
system-bath coupling strength $\gamma$ for the integrable spin-1/2 XXZ chain
with anisotropy (a) $\Delta = 1.5$ and (b) $\Delta = 3.0$, as resulting for
weak driving $\mu = 0.1$. Analogously, in (c) diffusion constants for the nonintegrable system with $\Delta'=0.5$ and $\Delta =1.5$ are depicted. Numerical data from exact stochastic unraveling (SU) are shown for two system sizes $L = 10, 20$. Numerical data from TEBD calculations up to system sizes $L=50$ are also indicated. For comparison, numerical data for the closed-system diffusion constant according to the Kubo
formula are depicted in (a) \cite{Karrasch2014}, (b) \cite{kraft2024}, and (c)
\cite{Wang2024}. Further, open-system diffusion constants $D_r(t)$ extracted at (representative) finite times $tJ=20$ from Figs.~\ref{Fig5}, \ref{Fig6} are indicated in purple.}
\label{Fig4}
\end{figure}
%------------------------------------------------------------------------------
%
%------------------------------------------------------------------------------
% numerical methods
%------------------------------------------------------------------------------
\subsection{Extracting diffusion constants from closed systems}

In closed quantum systems, we resort to LRT for the computation of the diffusion constant. In this context, a major role is played by the magnetization current $\mathcal{J}_{S}=\sum_{r} j_r$ (subscript `$S$' in $\mathcal{J}$ stands for the spin) and the corresponding autocorrelation function,
\begin{equation}
\langle \mathcal{J}_{S} \mathcal{J}_{S}(t) \rangle = \frac{\mathrm{Tr}\left[e^{-\beta H} e^{iHt} \mathcal{J}_{S} e^{-iHt} \mathcal{J}_{S}\right]}{Z} , \quad Z = \mathrm{Tr}\left[e^{-\beta H}\right] ,
\label{eq: autocorr_LRT}
\end{equation}
where $\beta = 1/T$ is the inverse temperature. As for the open-system scenario, we focus on the infinite-temperature regime, i.e.~$\beta \rightarrow 0$, such that Eq.\,(\ref{eq: autocorr_LRT}) reduces to 
\begin{equation}
\langle \mathcal{J}_{S} \mathcal{J}_{S}(t) \rangle_{\beta \rightarrow 0} = \frac{\mathrm{Tr}\left[ e^{iHt} \mathcal{J}_{S} e^{-iHt} \mathcal{J}_{S}\right]}{\text{dim }(H)}.
\label{eq:autocorr_LRT_reduced}
\end{equation}
where $\text{dim}$ denotes the full Hilbert-space dimension.
To study diffusion constants, we define the quantity~\cite{Steinigeweg2009}
\begin{equation}
\mathcal{D}_{\text{closed}}(t) = \frac{1}{\chi_s} \int_0^t dt' \, \langle \mathcal{J}_{S}(t') \mathcal{J}_{S} \rangle ,
\label{eq:D_of_t}
\end{equation}
where $\chi_s =1/4$ is the static susceptibility in the limit $\beta \rightarrow
0$.
The dc diffusion constant is then defined as $\mathcal{D}_{\text{dc}} =
\lim_{t \rightarrow \infty} \lim_{L \rightarrow \infty}
\mathcal{D}_{\text{closed}}(t)$, where the correct order of limits is crucial.
In practice, due to numerical constraints, calculations are limited to finite
systems. Consequently, a finite-size scaling is unavoidable for a reliable
estimation of the thermodynamic limit. In this work, we mostly rely on existing
results from the recursion method (RM)~\cite{Wang2024}, which allows for direct
evaluation of $\mathcal{D}_{\text{dc}}$ in the thermodynamic limit by
exploiting Lanczos coefficients in Liouville space.

\section{Numerical Methods}\label{sec:methods}
As already mentioned earlier, solving the GKSL master equation numerically is a formidable task since the complexity grows as $d^{2L}$, where $d$ denotes the local Hilbert-space dimension. We resort to two powerful state-of-the-art numerical techniques to find well-controlled approximated solution of the GKSL master equations as discussed below.
%------------------------------------------------------------------------------
% stochastic unraveling 
%------------------------------------------------------------------------------
\subsection{Stochastic unraveling}
The method of stochastic unraveling (SU), often also dubbed as Monte Carlo wavefunction technique, is a powerful technique to get a good approximate solution of the GKSL master equation. Like any Monte Carlo method, here the results are exact in the limit $N_{\text{sample}}\rightarrow \infty$ and we will shortly see the meaning of `sample' for this method.
\par 
Stochastic unraveling essentially aims at propagating pure quantum states $|\psi\rangle$ instead of
density matrices~\cite{Breuer2007, Dalibard1992, Michel2008, Daley04032014}, leading to 
different trajectories. 
In the end, the full density matrix is constructed by the average over these 
individual trajectories forming a stochastic solution of the GKSL master equation.
The approach consists of an 
alternating scheme: A deterministic time evolution generated by an effective
non-Hermitian Hamiltonian $H_\mathrm{{eff}}$, interspersed with stochastic
quantum jumps with one of the Lindblad operators at specific times.
\\
The deterministic time evolution is done with respect to the effective 
Hamiltonian
\begin{equation}
    H_{\mathrm{eff}} = H - \frac{i}{2} \sum_j \alpha_j L_j^\dagger L_j. 
\end{equation}
For our specific choice of Lindblad operators, cf.\ Eq.\,(\ref{eq: Lindblad 
operators}), the effective Hamiltonian can be rewritten as
\begin{equation}
    H_{\mathrm{eff}} = H - i\gamma + i\gamma \mu (n_{B_1} - n_{B_2}),
\end{equation}
with the occupation number $n_r = S^+_r S^-_r = S^z_r + \frac{1}{2}$ and bath sites $B_1=1$ and $B_2=N$. 
\\
Since $H_{\mathrm{eff}}$ is non-Hermitian, the resulting dynamics is not 
unitary. Thus, the norm of the state is not conserved and decays over time. 
When $\| |\psi(t)\rangle \|^2$ falls below a randomly drawn threshold 
$\epsilon \in ]0,1]$, a stochastic jump with one of the Lindblad operators 
occurs. 
The resulting renormalized state reads
\begin{equation}
    |\psi'(t)\rangle = \frac{L_j |\psi(t)\rangle}{\| L_j |\psi(t)\rangle \|},
\end{equation}
while the probability for each operator $L_j$ to be selected is given by
\begin{equation}
    p_j = \frac{\alpha_j \| L_j |\psi(t)\rangle \|^2}{\sum_j \alpha_j \| L_j |
    \psi(t)\rangle \|^2}.
\end{equation}
The deterministic evolution with $H_{\mathrm{eff}}$ then resumes. 
This process of alternating deterministic and stochastic steps generates a
specific trajectory $|\psi_T(t)\rangle$. Eventually, the average over a 
sufficiently large number of trajectories yields an arbitrarily good 
approximation of the time-dependent density matrix, thus yielding the solution of Eq.~(\ref{eq:lindblad}).
\\
Following this scheme, the local magnetization in Eq.~(\ref{eq: local magnetization})
can be rewritten as
\begin{equation}
    \langle S^z_r(t) \rangle \approx \frac{1}{N_{\text{traj}}} \sum_{k=1}
    ^{N_{\text{traj}}}
    \frac{\langle \psi_k(t) | S^z_r | \psi_k(t)\rangle}{\| \psi_k(t)\|^2}\ ,
\end{equation}
where $N_{\text{traj}}$ is the total number of trajectories. The same applies to
the local current in Eq.~(\ref{eq: local current}) which takes the form
\begin{equation}
    \langle j_r(t) \rangle \approx \frac{1}{N_{\text{traj}}} \sum_{k=1}
    ^{N_{\text{traj}}}
    \frac{\langle \psi_k(t) | j_r | \psi_k(t)\rangle}{\| \psi_k(t)\|^2}\ .
\end{equation}
 For $N_{\text{traj}}\to \infty$, this approximation becomes exact, and the 
 stochastic unraveling yields, in principle, an exact numerical solution of the
 GKSL master equation. Note that $N_{\text{sample}}$ in an usual Monte Carlo simulation in this case is analogous to $N_{\text{traj}}$.
\\

%------------------------------------------------------------------------------
\begin{figure}[t]
\includegraphics[width=1.\columnwidth]{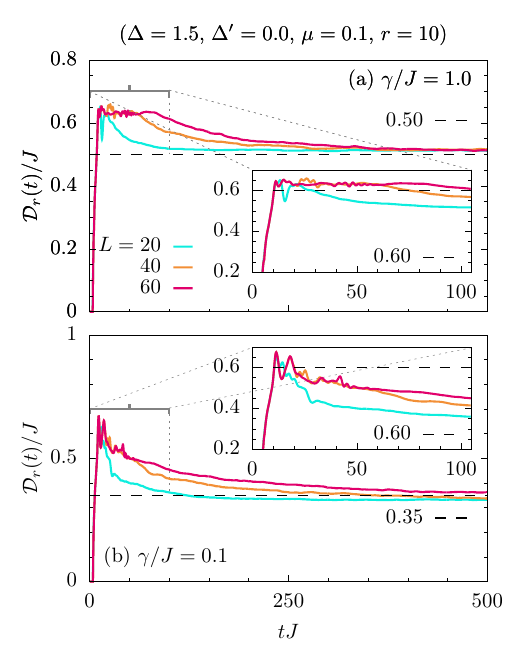}
\caption{Time dependence of local diffusion constant $\mathcal{D}_r(t)$ according to Eq.\,(\ref{eq: diffusion_constant}) as obtained from TEBD calculations for system-bath coupling (a) $\gamma/J=1.0$ and (b) $\gamma/J=0.1$. Data for site $r=10$ is depicted for system sizes $L=20,40,60$. Different time scales (short/long) are considered. The dashed lines in the main panels (a) and (b) correspond to the value of $\mathcal{D}_{\text{dc}}$ (cf.~Eq.\,(\ref{eq: diffusion constant from NESS})), while the dashed lines in the insets correspond to the Kubo value \cite{Wang2024}.}
\label{Fig5}
\end{figure}
%------------------------------------------------------------------------------
As mentioned before, we choose the initial condition $\rho(0) \propto
\mathbb{1}$, corresponding to the infinite-temperature limit $\beta \rightarrow
0$. This is reflected in the stochastic unraveling approach as a random pure 
initial state
\begin{equation}
    |\Psi(0)\rangle \propto \sum_{n=1}^{\text{dim}(H)} c_j |j\rangle ,\ 
\end{equation}
where $c_j=a_j+\mathrm{i}b_j$. 
The coefficients $a_j$ and $b_j$ are independent random numbers drawn from a 
normal distribution with zero mean and variance $1/2$. 

%------------------------------------------------------------------------------
% TEBD
%------------------------------------------------------------------------------
\subsection{Matrix product state (MPS) based technique}
As an alternative to stochastic unraveling, we resort to matrix product state (MPS) based technique to solve the GKSL master equation. In particular, we resort to time-evolving block decimation method (TEBD) \cite{Vidal_TEBD} to time evolve the vectorized density matrix $\rho_{\text{vec}}(t)$ written as a MPS \cite{Znidaric_PRL_2011, Bertini_rev, SN_superdiffusion, SN_stark}. More precisely, the GKSL master Eq.\,(\ref{eq:lindblad}) can be recast in the vectorized form as
\begin{equation}
    \partial_{t} \rho_{\text{vec}}(t) = \mathcal{L}_{\text{sup}} \rho_{\text{vec}}(t)\ ,
\end{equation}
where $\mathcal{L}_{\text{sup}}$ stands for the superoperator representation of the Liouvillian having the form $\mathcal{L}_{\text{sup}} = -i(H \otimes \mathbb{1} - \mathbb{1} \otimes H^{T}) + \sum_{j} \frac{\alpha{j}}{2} ( 2L_{j} \otimes L_{j}^{*} - L_{j}^{\dagger}L_{j} \otimes \mathbb{1} - \mathbb{1} \otimes L_{j}^{*}L_{j}^{T})$\,, where superscript $T$ and $*$ denotes transpose and complex conjugation respectively. Note that $\mathbb{1}$ is the identity matrix with dimension $\text{dim}(H)$.
For our TEBD calculation, we use bond dimension $\chi=150$ and time step $dt=0.2$, we have carefully checked that all our results are well converged both w.r.t bond dimension and time step. The TEBD is implemented using the ITensor (Julia) library \cite{Itensor_1}.

%------------------------------------------------------------------------------
% transport coefficients in isolated quantum systems
%------------------------------------------------------------------------------
%
%------------------------------------------------------------------------------
% results
%------------------------------------------------------------------------------
\section{Results}
\label{sec: results}
To make our presentation clear and compact, we first discuss the results for the integrable system followed by the results for the non-integrable one.
%------------------------------------------------------------------------------
% results integrable system
%------------------------------------------------------------------------------
\subsection{Integrable system}
For this subsection, we consider the spin-1/2 XXZ chain with open boundary conditions described by Eq.\,(\ref{eq: integrable system}).
In Fig.~\ref{Fig2} the real-time dynamics of (a) the local magnetization density
$\langle S_{r}^{z} \rangle$ and (b) the local current $\langle j_{r}\rangle$ for
a spin-$1/2$ XXZ chain with $L=20$ sites and anisotropy $\Delta=1.5$ under a
strong system-bath coupling $\gamma/J = 1.0$ and weak driving $\mu = 0.1$ is
shown. Data obtained by stochastic unraveling (SU) is averaged over
$N_{\text{traj}}=50000$ trajectories. The results
of both methods, SU and TEBD, show a reasonable agreement. 
While the remaining deviation stems from the approximative nature of the two methods
(finite number of trajectories/finite bond dimension), this deviation does not
affect the conclusions drawn later. As can be seen from Fig.~\ref{Fig2}, panel (a)
and (b), at times $t\gtrsim 100$, the resulting NESS is reached. The NESS
features a linear magnetization profile in the bulk of the system and a NESS
current independent of site location $r$. As a concrete example, in
Fig.~\ref{Fig2}(c), we show the corresponding magnetization profile for a chosen
time $tJ=200$. By fitting over the bulk sites $r=5\dots15$, we can extract a
linear slope of $\nabla S_{\text{ness}}^{z}=-0.0421 \mu/2$, which is also
consistent with the TEBD calculations.
In Fig.~\ref{Fig2}(b) the local current according to Eq.\,(\ref{eq: local current}) for the bulk sites $r=5\dots15$ is shown. As for the magnetization profile, also here the choice of the initial infinite-temperature state is reflected in $\langle j_r(t=0)\rangle=0$. For times $tJ\gtrsim 100$ the NESS is reached exhibiting a spatially homogeneous current with $j_{\text{ness}}=0.0225 \mu J/2$. 
\par
Analogous data is presented in Fig.~\ref{Fig3} for the case of a smaller system-bath coupling $\gamma/J = 0.1$.
Here, the local density in (a) as well as the local current in (b) reach the NESS at later times $tJ \gtrsim 150$ with $\nabla S_{\text{ness}}^{z}=-0.0397 \mu /2$ and $j_{\text{ness}}=0.0136 \mu J/2$. Note that for this case the relaxation time scale to the NESS is larger than for the previous case due to the weaker coupling to the bath. Once again, we obtain a very reasonable agreement of data from SU and TEBD calculations.
\\
%
%------------------------------------------------------------------------------
\begin{figure}[t]
\includegraphics[width=1.\columnwidth]{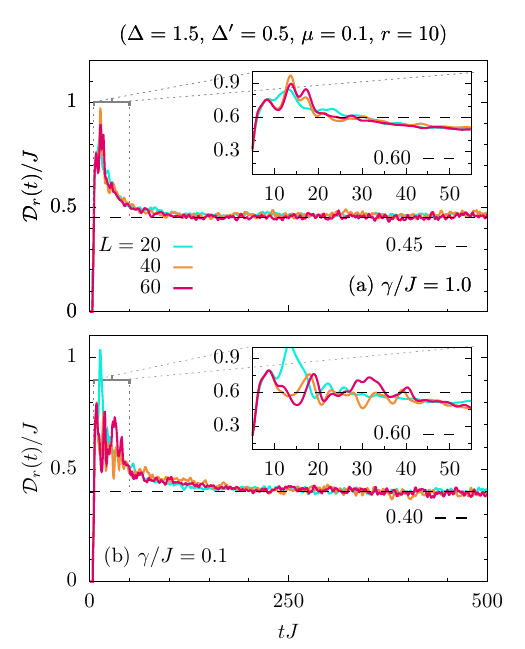}
\caption{Analogous data as depicted in Fig.~\ref{Fig5}, but now for the nonintegrable system with $\Delta' = 0.5$ according to Eq.\,(\ref{eq: nonintegrable system}).}
\label{Fig6}
\end{figure}
%------------------------------------------------------------------------------
%
Fig.~\ref{Fig4} shows the dc diffusion constant $\mathcal{D}_{\text{dc}}$ defined in Eq.\,(\ref{eq: diffusion constant from NESS}) for various system-bath coupling strengths $\gamma$ computed at $\mu=0.1$ with anisotropy (a) $\Delta=1.5$ and (b) $\Delta = 3.0$. 
Numerical data from SU is shown for two system sizes $L = 10$ and $L = 20$ and
is complemented by TEBD calculations for larger system sizes up to
$L=50$. For comparison, numerical data for the closed-system diffusion constant
obtained via Kubo formula are also presented \cite{Karrasch2014, Wang2024,
kraft2024}. Clearly, a strong $\gamma$ dependence is visible in the
$\mathcal{D}_{\text{dc}}$ for both values of anisotropy and all system sizes
considered. Since the dc diffusion constant is a bulk material property, it is
physically expected to be independent of specific details such as the
system-bath coupling strength $\gamma$ and system size $L$. Thus,
the (strong) variation of $\mathcal{D}_{dc}$ with system-bath coupling strength
$\gamma$ clearly indicates a limitation in accessing the dc diffusion constant
from the open-system method. This constitutes a central result of our work.

To understand the origin of the aforementioned limitation and to find
a practical solution, we use the open system based method to probe the dynamical
diffusion constant $\mathcal{D}_{r}(t)$, defined in Eq.\,(\ref{eq:
diffusion_constant}), for a representative and fixed site
$r=10$, which is not changed with $L$ and sufficiently away from the bath
contact. Fixing $r$ allows for a convergence analysis with respect to time,
which is presented in Fig.~\ref{Fig5} for two different values
of $\gamma/J\in\{0.1, 1.0\}$ and fixed $\mu=0.1$. The insets in both panels show a
zoomed-in view of the plateau region of the $\mathcal{D}_r(t)$ curve,
highlighted by the rectangular boxes. The value of the diffusion constant for the
representative site $r=10$ and $L=60$ within this plateau region at a chosen time $tJ=20$
agrees quite reasonably with the value of the diffusion constant
extracted from the Kubo formula, as shown in the Fig.~\ref{Fig4} (see the violet `$\times$'
points in panel (a)). Now, few important comments are in order:
i) As can be seen in Fig.~\ref{Fig4}, these short-time results of the diffusion constant extracted from the plateau has minimal $\gamma$ dependence and as mentioned earlier, share reasonable agreement with the results from the Kubo formalism. This leads us to consider the $\gamma$ dependence of transport coefficients as a potential diagnostic of finite-size effects. 
ii) The diffusion constant only has a very weak time dependence within the plateau. Therefore, although changing the reference time from $tJ=20$ to some other time within the plateau would change the value of the time-dependent diffusion constant, but that change will be rather small and hence will not affect the main conclusions for all practical purposes. On a similar note, we may point out that for $\gamma/J=0.1$, strictly speaking, $tJ=20$ is not within the `plateau' for $L=20$, but the deviation from the former is also rather small and hence does not alter our main conclusions. iii) Followed by the plateau, the $\mathcal{D}_r(t)$ curve takes a downward turn and finally reaches the NESS. 
This downward turn can possibly be a signature of the onset of finite-size effects. 
iv) The oscillation within the plateau window emanates from the fact that TEBD is only an approximate method and additionally, the data for smaller $\gamma$ ($\gamma/J=0.1$ for example) are more `error prone'. This is a consequence of a weaker coupling to the baths, which is often not sufficient to suppress the operator state entanglement entropy growth (a measure of simulability) but necessary for MPS-based methods to work well \cite{Alba2022}. Indeed we have more stable data for $\gamma/J=1.0$.
v) The onset of the plateau region essentially depends on location $r$ and hence to make the comparison between different $L$ data on a same footing, we fix the reference site. vi) As we see in the insets, the width of the plateau depends on system size $L$ and \textit{roughly} scales as $\sim L$. In contrast, in a typical diffusive system, or more precisely, in the diffusive spin-$1/2$ XXZ chain, the NESS time scale behaves as $L^{2}$. This explains why the NESS results calculated from open-system approaches, such as $\mathcal{D}_{\text{dc}}$, are contaminated by finite-size effects.

%------------------------------------------------------------------------------
% results nonintegrable system
%------------------------------------------------------------------------------
\subsection{Nonintegrable system}
\label{sec: nonintegrable system}
Proceeding further, we consider the nonintegrable spin-$1/2$ XXZ chain described by Eq.~(\ref{eq: nonintegrable system}) with $\Delta'=0.5$ denoting the strength of the integrability-breaking term. 
Essentially, we redo the numerical calculations in Fig.~\ref{Fig2} and Fig.~\ref{Fig5} to study the impact of integrability breaking. The time evolutions of the local current $j_r(t)$ and local magnetization $S_r^z(t)$ can be found in App.~\ref{appsec: Local density and current in the nonintegrable system}. 
In Fig.~\ref{Fig4} (c) diffusion constants extracted in the long-time limit according to Eq.\,(\ref{eq: diffusion constant from NESS}) are presented. Also here, we find a clear and strong $\gamma$-dependence of the diffusion constants for all system sizes considered consistent with SU as well as TEBD calculations.
In Fig.~\ref{Fig6} the time-dependent local diffusion constant $\mathcal{D}_{r}(t)$ at a chosen site $r=10$ is depicted analogously to Fig.~\ref{Fig5}.
While the plateau width is shorter compared to the integrable case, somewhat similar conclusions follow. Once again, for the chosen representative site and a chosen time point $tJ=20$ within the plateau, we find the time-dependent diffusion constant to be close to the ones obtained from the Kubo formalism and has minimal $\gamma$ dependence. 

%------------------------------------------------------------------------------
\begin{figure}[t]
\includegraphics[width=0.85\columnwidth]{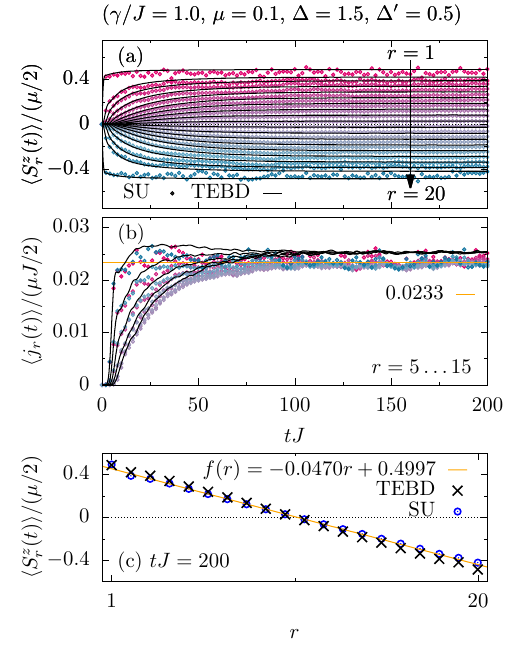}
\caption{ Analogous data to Fig.~\ref{Fig2} now for the nonintegrable spin-1/2 XXZ chain
with next-nearest neighbor interaction $\Delta'=0.5$ according to Eq.\,(\ref{eq: nonintegrable system}).}
\label{Fig7}
\end{figure}
%------------------------------------------------------------------------------

%------------------------------------------------------------------------------
% conclusion
%------------------------------------------------------------------------------
\section{conclusion and future outlook}   
\label{sec: conclusions}
To summarize, we have revisited the method of boundary driving via Lindblad baths employed for studying high-temperature spin transport and as a concrete example we considered the well-studied spin-1/2 XXZ model for our purpose. The primary goal was to understand the equivalence between state-of-the-art methods for computing transport coefficients—specifically, those based on closed-system dynamics such as linear response theory (LRT) and the Kubo formalism, and those based on boundary-driven open quantum systems. While certain dynamical aspects have recently been shown to be equivalent in these two seemingly distinct approaches, their correspondence at the level of transport coefficients remains an open and challenging question.
\par
To this end, we employed advanced numerical techniques such as stochastic
unraveling and matrix-product-state based algorithms to extract diffusion
coefficients via open-system dynamics by solving the GKSL master equation. Our
numerical results for dc diffusion constant
$\mathcal{D}_{\text{dc}}$, extracted from the NESS, showed a mismatch with
those obtained from the closed-system dynamics based on the linear-response Kubo
formalism. Additionally, we observed a strong dependence of
$\mathcal{D}_{\text{dc}}$ on the system-bath coupling strength $\gamma$, a
phenomenon that shows up irrespective of system size considered in our work
(ranging from $L=10$ to $L=60$). Such a dependence of a material constant like
$\mathcal{D}_{\text{dc}}$ on details of system-bath coupling is unphysical.
Therefore, our results indicate that the dc diffusion constant computed from
boundary-driven methods are not reliable.

To find out the source of such a
problem in open-system techniques and also to find a reasonable resolution, we
computed the corresponding time-dependent diffusion constants $\mathcal{D}(t)$, 
where our key findings are sketched in Fig.~\ref{sketch}. 
In the long-time limit we found: (i) The dc diffusion constant $\mathcal{D}_{\text{dc}}$ extracted from the boundary-driven approach disagrees with the closed-system Kubo-based result. (ii) In the open-system setup, $\mathcal{D}_{\text{dc}}$ exhibits a strong dependence on the system-bath coupling strength $\gamma$. To identify the origin of this mismatch, we analyzed the full time evolution of $\mathcal{D}(t)$ within the boundary-driven framework and and further compare it to the Kubo-based closed-system result. Here, we found for (earlier) times $tJ \leq t_{\mathrm{fs}}J$, the open-system diffusion constant $\mathcal{D}(t)/J$ converges to a plateau value which shows a convincing agreement with the Kubo-based closed-system result and is nearly independent of system-bath coupling $\gamma$.

We suggest, that a physical interpretation follows from the finite-size time scales: For our model, the onset of finite-size effects scales as $t_{\mathrm{fs}}J \propto L$ \cite{Bertini_rev}. Increasing $L$ therefore extends the initial plateau in $\mathcal{D}(t)$ and prolongs its agreement with the closed-system value, largely independent of $\gamma$. However, the approach to the NESS occurs on a (later) timescale $t^{*}J \propto L^{2}$, so NESS-based dc values from open-system approaches (such as $\mathcal{D}_{\text{dc}}$) remain affected by finite-size effects and not agreeing with the closed-system Kubo-based value. 
In other words, boundary-driven schemes effectively realize the order of limits
$\lim_{L \to \infty}\lim_{t \to \infty}$, in contrast to $\lim_{t \to
\infty}\lim_{L \to \infty}$. Since these limits generally do not commute,
NESS-based results for $\mathcal{D}_{\text{dc}}$ at finite $L$ are contaminated
by finite-size effects. This also allows for the interpretation of the
sensitivity of $\mathcal{D}_{\text{dc}}$ to $\gamma$ from open-system approaches
as a diagnostic of finite-size effects.

We emphasize that, despite limitations in computing
long-time transport coefficients, the boundary-driven open system approach
remains highly valuable. Beyond enabling the extraction of accurate
time-dependent transport coefficients, as demonstrated in this work, it provides
a powerful framework to probe transport behavior via the scaling of the
steady-state current $j_{\text{ness}}$ with system size
$L$. Compelling numerical evidence from previous studies \cite{SN_stark,
SN_superdiffusion, SN_diffusion, Peter_stark} showed  that this method
efficiently captures not only diffusive transport but also a wide range of
anomalous regimes, including superdiffusion and fractonic subdiffusion.
\par
As a future vision, further investigations, extending to other open systems and
also considering other boundary conditions as well as other driving strengths
are promising future directions for a deeper understanding of the connection
between open and closed quantum systems with respect to transport phenomena.
Particularly interesting is the question whether or not
the width of the plateau in $\mathcal{D}_r(t)$ scales faster than $L^2$ for a
class of generic nonintegrable systems.
%------------------------------------------------------------------------------
\begin{figure}[t]
\includegraphics[width=0.85\columnwidth]{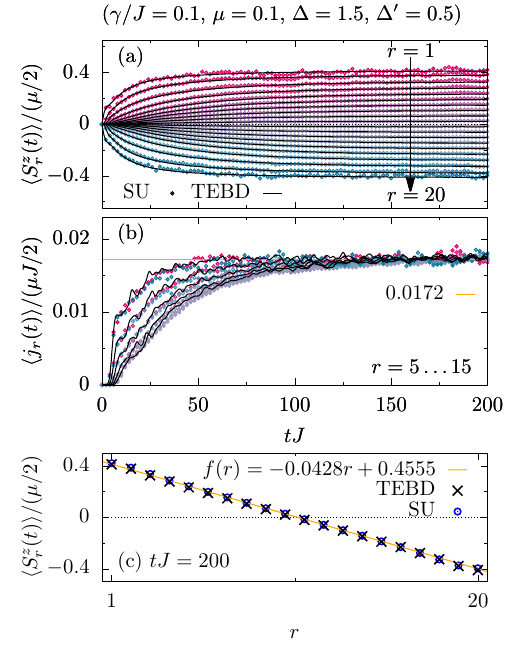}
\caption{Analogous data to Fig.~\ref{Fig7} now for a system-bath coupling $\gamma/J=0.1$.}
\label{Fig8}
\end{figure}
\\
%------------------------------------------------------------------------------
%-------------------------------------------------------------------------------
% Acknowledgments
%-------------------------------------------------------------------------------
\section*{Acknowledgments}
We thank Zala Lenarčič for insightful discussions. This
work was funded by the Deutsche Forschungsgemeinschaft (DFG)
under Grants No. 397107022, No. 397067869, and No. 397082825 within the DFG
Research Unit FOR 2692 under Grant No. 355031190.
Additionally, we greatly acknowledge computing time
on the HPC3 at the University of Osnabrück, granted by
the DFG, under Grant No. 456666331. All the tensor-network based calculations
were performed at the HPC
cluster facility at MPI-PKS Dresden. 
%-------------------------------------------------------------------------------
% Data availability
%-------------------------------------------------------------------------------

\section*{Data availability}
The data that support the findings of this article are openly
available at \cite{dataset} .

%-------------------------------------------------------------------------------
% Appendix
%-------------------------------------------------------------------------------
\appendix
%-------------------------------------------------------------------------------
% Appendix: results nonintegrable system
%-------------------------------------------------------------------------------
\section{Local density and current in the nonintegrable system}
\label{appsec: Local density and current in the nonintegrable system}
To extract the long-time limit diffusion constants according to Eq.\,(\ref{eq: diffusion constant from NESS}), we recalculate the time evolution in the open system for the nonintegrable spin-$1/2$ XXZ chain with $\Delta=1.5$ and $L=20$ sites analogously to Fig.~\ref{Fig2} and Fig.~\ref{Fig3}. Corresponding results for $\gamma/J=1.0$ and $\gamma/J=0.1$ are depicted in Fig.~\ref{Fig7} and Fig.~\ref{Fig8}, respectively. For both values of $\gamma$, we find the NESS to set in approximately at times $tJ\gtrsim 100$ in accord with the results for the integrable system. Note that the order of magnitude for both, local magnetization as well as local current, is very similar to corresponding values of the integrable system. Also here, we observe a remarkable agreement of SU and TEBD calculations.
%\bibliographystyle{apsrev4-1_titles}
%\bibliography{notes.bib}

\clearpage
\newpage
\end{document}